\documentclass[aps,prb,amsmath,twocolumn,amssymb,floatfix,showpacs,nofootinbib,longbibliography]{revtex4-2}
\usepackage{MnSymbol}
\usepackage{braket}
\usepackage[dvipsnames]{xcolor}
\usepackage{float}
\usepackage{subfigure}
\usepackage[colorlinks=true,linktoc=page,linkcolor=blue,citecolor=blue,urlcolor=blue]{hyperref}
\usepackage{multirow}
\usepackage[T1]{fontenc}
\usepackage[utf8]{inputenc}
\usepackage[normalem]{ulem}
\usepackage{sidecap,tikz}
\usepackage{mathrsfs}
\usepackage{amsmath}

\usepackage{ulem}  

\mathchardef\mhyphen="2D 

\definecolor{lime}{HTML}{A6CE39}
\DeclareRobustCommand{\orcidicon}{\hspace{-1.0mm}
    \begin{tikzpicture}
        \draw[lime, fill=lime] (0.0,0.0) 
        circle [radius=0.15] 
        node[white] {{\fontfamily{qag}\selectfont \tiny \,ID}};
        \draw[white, fill=white] (-0.0525,0.095) 
        circle [radius=0.007];
    \end{tikzpicture}
    \hspace{-3.0mm}
}
\foreach \x in {A, ..., Z}{\expandafter\xdef\csname orcid\x\endcsname{\noexpand\href{https://orcid.org/\csname orcidauthor\x\endcsname}{\noexpand\orcidicon}}
}

\allowdisplaybreaks

\begin{document}
\title{Temperature and magnetic field dependent $g$-factors in electron spin resonance spectroscopy}  

\author{Sebastian Kalhöfer\orcidA{}}
\email{sebastian.kalhofer@physics.uu.se}
\affiliation{Department of Physics and Astronomy, Uppsala University, Box 516, 75120 Uppsala, Sweden}

\begin{abstract}
We propose that phonons coupled to electrons play an important role in both the fundamental phenomenology of electron spin resonance as well as the interpretation of such measurements. By including spin-dependent electron-phonon interactions originating from spin-orbit coupling, we demonstrate that the g-factor acquires emergent temperature dependencies that are in excellent agreement with experimental observations. Moreover, we predict that at low temperatures, the $g$-factor will exhibit a non-trivial dependence on the magnetic field, indicating a dynamic property that has not been previously considered in electron-phonon systems. We further predict the influence of phonons to be negligible at very low and very high temperatures, with the former requiring experimental verification.
\end{abstract}

\maketitle

\section{Introduction} \label{introduction}
Electron spin resonance (ESR), also known as electron paramagnetic resonance (EPR), is a powerful spectroscopic technique that has found widespread usage across technological applications and diverse scientific fields, including physics, chemistry, materials science and biology \cite{59_Freed2000, 37_ESR12, 50_Griller1980, 51_Goodman1970, 52_Sahu2013-mv, 53_J_R_Mallard1969-iw, 54_Bennati_2005}. ESR has been harnessed for various applications, for example magnetic resonance imaging (MRI) in medicine, but also for the development of advanced materials for electronics and spintronics devices \cite{58_Bader2010,60_Awschalom2013,55_Sanvito2011-gp,56_LEE2008}.
At the heart of ESR lies the interaction between electronic spins in a paramagnetic sample with an external magnetic field, 
such that the absorption of energy (i.e. excitation of states) at specific resonance frequencies becomes possible. 
This basic principle results in sample specific ESR spectra that provide a wealth of information about the sample's electronic structure 
\cite{20_Bolvin2015, 23_Rieger2007-ru, 39_ESR01, 30_ESR03, 31_ESR04, 32_ESR05, 33_ESR06, 34_ESR09, 35_ESR10, 36_ESR11, 37_ESR12}.
Despite the impressive achievements of ESR, there are still significant shortcomings in the current theoretical framework, which are sometimes qualitatively discussed but hardly quantitatively examined, namely the use of Hamiltonians that are formulated as effective single particle models at absolute zero temperature \cite{37_ESR12, 20_Bolvin2015, 23_Rieger2007-ru}. While these models have been invaluable in explaining many observed ESR spectra, they fall short in accounting for e.g. temperature dependencies and a typical rise of the $g$-factor at low temperatures. Experimental studies attribute the observed rise in the g-factor at low temperatures to a magnetic phase transition or electron-electron interactions. In our model, we independently identify a magnetic phase transition, providing a consistent theoretical framework for this observation.
One particularly notable approximation is neglecting electron-phonon interactions in ESR systems in the solid state, which have long been treated as not operating on the spin degrees of freedom in model Hamiltonians. Electron-phonon systems refer to systems in which the coupling between electronic degrees of freedom and quantized lattice vibrations (phonons) significantly affects observable properties. This interaction is central to various phenomena, including electrical resistivity, superconductivity \cite{24_PONCE2016116, 25_Giustino2007, 01_RevModPhys.89.015003, 06_Mahan2012, 05_Marsiglio2020, 02_PhysRevB.87.024505}, and, as this work suggests, the response in ESR. In essence, the only role that phonons acquire in traditional ESR-spectroscopy, is that of a relaxation-channel for the spins \cite{36_ESR11}. Whereas this approximation is widely used \cite{01_RevModPhys.89.015003, 48_PhysRevLett.108.167402, 02_PhysRevB.87.024505, 05_Marsiglio2020} and makes many calculations much easier, there is an increasing amount of scientific evidence, in particular coming from studies of the chiral induced spin selectivity (CISS) effect, that indicates the significance of spin dependent electron-phonon interactions \cite{27_PhysRevB.102.235416, 26_Fransson2019-py, 19_PhysRevResearch.5.L022039}.
In this article, we aim to shine light on the shortcomings of the existing ESR theory for solid state systems as for example electrons in a crystal, particularly the role of temperature dependence, and propose a fresh perspective. We assert that the inclusion of spin dependent electron-phonon interactions arising from the natural quantization procedure of the relativistic spin-orbit interaction can open new ways for understanding spin systems in external magnetic fields. These theoretical insights may have far reaching implications by helping in the identification of crucial interactions governing the level structure of spin systems under ESR conditions.
In this work we show how the interaction to a phonon bath severely influences the energy spectrum by equipping the energy levels with a temperature dependence which leads to a non-linear dependence on the magnetic field strength. The latter represents an effect which implies that the measured spectrum, and the conclusions that are drawn about g-factors, depend on the frequency of the photons, that are used to excite the electrons of the sample. To our knowledge, this effect is only known in heavy-fermion compounds \cite{40_30_Referee_Article, 41_31_Referee_Article, 42_32_Referee_Article}. Our study reveals, that these dependencies can also occur in electron-phonon systems and therefore constitutes a crucial ingredient for the prediction and not just reproduction of ESR-spectra. Lastly, the Green-function formalism that our study uses can easily be generalized to heavy-fermion compounds, making it a viable guideway to quantitatively study these systems.
\section{Theory} \label{Sec:drivenchern}
For arbitrary temperatures, it is common \cite{20_Bolvin2015} to invoke the concept of a pseudo spin ${\bf S}$ instead of the actual electron spin, and define a phenomenological pseudo spin Hamiltonian which reads to lowest order
\begin{align}
\label{PhenomenologicalHamiltonian}
H =&
	\beta_e{\bf B}\cdot\mathbb{G}\cdot {\bf S}
	+
	{\bf I}\cdot\mathbb{A}\cdot{\bf S}
	+
	{\bf S}\cdot\mathbb{D}\cdot{\bf S}
\end{align}
for a system in a magnetic field ${\bf B}$ with $3 \times 3$-matrices $\mathbb{G}$, $\mathbb{A}$, $\mathbb{D}$ and nuclear spin ${\bf I}$. 
This ansatz relies on the assumption that there is a correspondence between the pseudo spin ${\bf S}$ and true eigenstates.
The term on the left, in the middle and on the right represent the Zeeman splitting, hyperfine coupling interaction and the zero field splitting respectively. 
The values of $S$ (corresponding to its degeneracy $2 S + 1$), ${\bf I}$ and the matrices parametrizing the various 
interactions are determined by requiring that the transitions between eigenstates reproduce the experimentally observed spectrum. 
There are also higher order contributions and further interactions like 
spin-orbit coupling which can be implemented into Eq. (\ref{PhenomenologicalHamiltonian}) in order to reproduce the 
experimental data, yielding additional sets of parameters. Influences from electron-phonon interactions have also been observed and investigated, 
but a description that quantitatively outlines how temperature effects arise and influence the energy levels seem to be 
hardly discussed so far \cite{49_PhysRevB.101.184104, 47_Bray2022, 08_PhysRev.166.307}. 
While traditionally one computes the spectrum of the Hamiltonian to determine excitation energies from differences in eigen-energies, 
our approach builds on this foundation by concentrating on the peaks of the electronic density of states of a coupled electron-phonon system to 
extract excitation energies. This parallels the methodology of traditional ESR, where differences in eigen-energies define g-factors. 
For proof of concept, we consider the simplest form of Eq. \eqref{PhenomenologicalHamiltonian}, that is in SI-units \cite{38_20_Nolting}
\begin{align}\label{Zeeman}
	H_0
	\equiv 
	- \boldsymbol{\mu} \cdot \boldsymbol{B} 
	\equiv 
	g_{ \text{e}} \frac{ | e | }{2 m_{\text{e}}} \boldsymbol{S} \cdot \boldsymbol{B}
\end{align}
with the electronic $g$-factor $g_e=2.002 $, the electron charge and mass $e, m_{\text{e}}$, the spin operator $\boldsymbol{S}$ and magnetic field vector ${\bf B}$. Notice that this is just the Zeeman term that appears in the Pauli equation for an electron.
In this paper we consider the minimal example of the spin system in Eq. (\ref{PhenomenologicalHamiltonian}), namely the spin $1/2$ system in Eq. (\ref{Zeeman}) and demonstrate how electron-phonon-interactions can influence the energy levels. Using physically reasonable numbers for the variables, we show in particular how temperature and magnetic field dependent $g$-factors can emerge.
The Hamiltonian in Eq. (\ref{Zeeman}) has eigenstates given by spin-up ($+$) and spin-down ($-$) states. The corresponding eigenvalues are linear functions in the magnetic field
\begin{align}\label{eq:non_interacting_energies}
E_{0, \pm} ( B ) 
=
\pm 
g_{ \text{e}} \frac{ |e| }{2 m_{\text{e}}} \frac{\displaystyle \hbar}{2} B 
=
\pm
\frac{ g_e \mu_B B}{2}
\end{align}
with the Bohr-magneton $\mu_B = \frac{ |e| \hbar }{2 m_e}$. The underlying density of states has two peaks and therefore, in order to excite an electron from the lower energy level to the upper by means of radiation, the photons must carry a frequency $\nu$ such that
\begin{align}\label{Eq:Classical_g_factor}
h \nu
= 
g_e
\mu_B
B
	.
\end{align}
In a typical ESR experiment, one fixes the frequency $\nu$ and measures the energy-loss to the sample when one changes the magnetic field. This equation is then inverted such that $g_e$ becomes a function of all the other quantities which are either assumed to be known, or measured experimentally. 
Similar to how ESR determines resonance frequencies by considering energy level separations under a magnetic field, we examine peak separations in the density of states, which directly correspond to excitation energies.

In this paper, we perform the following steps: First we specify the system by considering interactions ${\bf V}$ with a phononic bath. Then we determine the self-energy $\boldsymbol{ \Sigma }$ (and therefore the retarded Green function) of the interacting spin-system in second order perturbation theory. Subsequently we fix the number of electrons to half-filling, by numerically determining the temperature-dependent chemical potential. Finally, we determine the energy-difference $ E_{ + } - E_{ - } \equiv \Delta E \equiv \Delta E \, ( T , B, {\bf V})$ that one has to put into the system in order to excite the system from one spin-state to another as a function of the magnetic field and temperature. The effective $g$-factor is then given by the equation 
\begin{align}\label{gEffective}
g_{\text{eff}} (T, B, \boldsymbol{V}) := 
\frac{
\Delta E \, (T, B, {\boldsymbol{V}})}{\mu_B B}
	,
\end{align}
with an expression for $\Delta E \, (T, B, {\bf V})$ which is derived below.
Connection with experimental ESR measurements may be established by considering, for instance, localized orbitals or defects embedded in a crystalline environment \cite{10_PhysRev.75.1412, 14_gvsTref2}. In both these cases, the ESR driving field excites the state and due to spin-orbit coupling, the spin-components of the electronic orbitals may be coupled to the phonons. In this way, the underlying energy levels are dressed with a non-trivial temperature-dependence. In this sense, the electron-phonon interaction in this environment provides an effective spin-dependent coupling between the localized states and the phonons.
Recently, it was demonstrated \cite{27_PhysRevB.102.235416,19_PhysRevResearch.5.L022039} that a spin-dependent electron-phonon interaction emerges from the spin-orbit coupling
\begin{align}\label{Eq:Spin_Orbit_Coupling}
H_{\text{soc}} 
= 
\frac{\hslash}{4m^2 c^2} \boldsymbol{\sigma} \cdot\Bigl[ \boldsymbol{ \nabla } V(\boldsymbol{ r })\times \boldsymbol{ p } \Bigr]
.
\end{align}
The electron-phonon coupling arises by considering that the electronic confinement potential $V(\boldsymbol{ r })$ is build up by the interactions between the electrons and the nuclei. Since the nuclei are not fixed in space but move about their equilibrium positions $\boldsymbol{ R }_0$, the electron-phonon coupling is associated with the gradient of the potential around $\boldsymbol{ R }_0$. Considering the basis of localized orbitals or maximally localized Wannier-functions \cite{22_RevModPhys.84.1419, 24_PONCE2016116, 25_Giustino2007} one sees that single-site spin-dependent electron-phonon couplings are forbidden due to time reversal symmetry. The simplest model that does not suffer from this restriction are two degenerate electronic energy-levels with spin, giving four states in total. 
By the above argument, we consider a simple ESR system which includes spin-dependent electron-phonon-interactions due to spin-orbit coupling, but for the sake of the argument, we neglect the phonon-independent contribution of the spin-orbit coupling, which produces merely a relative shift and no temperature-dependence in electronic energies. We consider the Hamiltonian $H_\text{el}$ of two degenerate electronic orbitals that are in an external magnetic field $B$, a single phonon mode Hamiltonian $H_\text{ph}$ \cite{09_Iaru2021} corresponding to a contribution of, e.g., optical phonons at the $\Gamma$-point, and an interaction-Hamiltonian $H_\text{int}$, such that the total set-up is written as 
\begin{align}\label{Eq:Hamiltonian_tot}
H_{\text{tot}} = H_{\text{el}} + H_{\text{ph}} + H_{\text{int}}
	.
\end{align}
Concretely, the electronic orbitals and the free phonons are provided by $ H_\text{el} = \psi^\dagger \boldsymbol{ \epsilon } \psi $ and 
$H_\text{ph} = \hslash \omega b^\dagger b = \Omega b^{\dagger} b$, respectively, with spinors $\psi
=
(\psi_1 \, \psi_2)^t$, $\psi_n=(\psi_{n + } \, \psi_{ n - })^t$ whereas $n=1,2$ corresponds to the two degenerate electronic orbitals, and $\pm$ 
denotes the spin.
Notice that we deliberately neglect phonon-independent spin-orbit interactions in this model, which would play an integral role in more realistic spin systems. Their inclusion leads to an anisotropic modification of the energy levels, manifesting as zero-field splitting for more realistic spin-systems and a tensor structure of the $g$-factor, as in Equation (\ref{PhenomenologicalHamiltonian}). However, since these effects do not introduce an intrinsic temperature dependence, which is the focus of this article, we omit them for the sake of notational clarity.
Here, without loss of generality set the on-site energy to zero $\epsilon_0=0$, 
and therefore obtain the non-interacting Hamilton-matrix 
$ \boldsymbol{ \epsilon } = \operatorname{diag} ( \epsilon_z, -\epsilon_z, \epsilon_z, -\epsilon_z) $, 
such that the energy spectrum is given by $\pm \epsilon_z = \pm \frac{ g_e \mu_B B}{2}$.
For future reference we also denote the electronic energy-levels with respect to the chemical potential as $\omega_{\pm} = ( \pm \epsilon_z - \mu ) / \hslash$ and $\Omega_{\pm} = \hslash \omega_{ \pm }$. We assume a single phonon mode $\omega$.
By the above argument, we assign the interaction Hamiltonian $H_\text{int}$ as,
\begin{align}
H_{ \operatorname{int}}=&
	\sum_{mn}
		\psi^\dagger_m
		\boldsymbol{ V }_{mn}
		\psi_n
		\Bigl(
			b + b^{ \dagger }
		\Bigr)
\end{align}
where $ \boldsymbol{ V }_{12} =V_0 \sigma_0+\boldsymbol{ V }_1\cdot\boldsymbol{\sigma}$, $\boldsymbol{ V }_{21} = \boldsymbol{ V }_{12}^{\dagger}$, and $\boldsymbol{ V }_{11} = \boldsymbol{ V }_{22} = 0$ is the spin-dependent coupling between electrons and phonons, whereas the vanishing of the diagonal components $\boldsymbol{V}_{n n} = 0$ originates from time reversal symmetry \cite{27_PhysRevB.102.235416, 44_PhysRevB.92.100402, 26_Fransson2019-py, 19_PhysRevResearch.5.L022039}.
We connect the physics of this model to the single electron Green function $ \mathscr{G}( \mathrm{i} \omega_n )
=
\left( \mathrm{i} \hslash \omega_n - ( \boldsymbol{ \epsilon } + \boldsymbol{ \Sigma } ( \mathrm{i} \omega_n) - \mu ) \right)^{-1} $ from which we extract the average particle number 
\begin{equation}\label{Eq:Particle-Number}
\langle N \rangle
=
\frac{- \hbar }{ \pi } 
\int \mathrm{d} \omega
\,
n_{ F } ( \omega ) 
\operatorname{Im} 
\operatorname{Tr}
G^{ \, \text{ret}} ( \omega + \mathrm{i} 0 ),
\end{equation}
the spin resolved density of electron states, also called partial density of states,

\begin{equation}\label{eq:Spin-resolved-DOS}
\rho_\sigma(\omega)
=
- \hslash \operatorname{ Im } \sum_m G^{ \, \text{ret} }_{ \substack{m, m \\ \sigma ,\sigma } } (\omega + \mathrm{i} 0) / \pi 
\end{equation}

and the magnetization along the $z$-axis $M_z$ as 
\begin{equation}\label{Eq:Magnetization}
M_z
=
\frac{ g_e |e| \hbar^2 }{4 \pi m_e}
\sumint_{ \omega , j }
n_{F} ( \omega )
\operatorname{Im}
\left[
\left(
G_{ \substack{ j, j \\ \uparrow , \uparrow } }^{ \, \text{ret} } 
-
G_{ \substack{ j, j \\ \downarrow , \downarrow } }^{ \, \text{ret} } 
\right) ( \omega + \mathrm{i} 0 )
\right]
\end{equation}
where $G^{ \, \text{ret} }$ is the retarded form of the Green function, and we used the fluctuation-dissipation-theorem
\begin{equation}
G_{ \alpha \alpha^{\prime}}^{ < } ( \omega ) 
= 
- 2 \mathrm{i} \, n_{ F } ( \omega ) 
\operatorname{Im} 
G_{ \alpha \alpha^{\prime}}^{ \, \text{ret}} ( \omega + \mathrm{i} 0 ) ,
\end{equation}
with the lesser Green-function $ G_{ \alpha , \alpha^{ \prime } }^{ < } (t) 
= 
+ \frac{\mathrm{i}}{\hbar} 
\left\langle 
    c_{ \alpha^{ \prime }}^{ \dagger } ( 0 ) c_{ \alpha } (t) 
\right\rangle 
$ \cite{126_Nolting2012-ne, 127_Negele1998-al, 28_altland_simons_2010, 128_Bruus2004-ju, 06_Mahan2012}. Notice that the total density of states is given as the sum of the partial densities of respective spins $\rho = \rho_{ + } + \rho_{ - }$ and that a magnetic susceptibility can be defined as derivative $ \chi ( B , T ) \equiv \chi_{B} (T ) \equiv \left( \partial M_z / \partial B \right) ( B , T )$ \cite{121_Kardar2007-rw, 122_Schwabl2006-sb}. The chemical potential $\mu$ is determined numerically for each temperature, by using Equation (\ref{Eq:Particle-Number}) and  fixing the average number of particles to two, i.e. considering half-filling.
Considering the interactions with the phonons, the electronic single-particle Green function can be obtained from the Dyson equation $ \mathscr{G} = \mathscr{G}_0 + \mathscr{G}_{0} \Sigma \mathscr{G} $, where $\mathscr{G}_0( \mathrm{i} \omega_n ) 
= 
( \mathrm{i} \hbar \omega_n - ( \boldsymbol{ \epsilon } - \mu ) )^{-1}$ is the unperturbed Green function, and the self-energy ${\boldsymbol{ \Sigma }}$ includes the interactions between the electrons and phonons. The details concerning the self-energy are considered below. Before going into the details of the self-energy, we make some general remarks.
The electronic density of states that are associated to the non-interacting system, i.e. without phononic interactions,has two delta-peaks at the non-interacting energies $ \pm \epsilon_z \equiv E_{ 0 , \pm } $ from Equation (\ref{eq:non_interacting_energies}). The inclusion of interactions modifies this electronic density in Equation (\ref{eq:Spin-resolved-DOS}), by shifting number, location, broadening and height of the peaks. If we focus attention on the peaks that are are close to the non-interacting energy-levels (these will turn out to be the largest in height), we can associate average electronic energies $ E_{ \pm } $ for the interacting system with these peaks, which depend on temperature $T$, magnetic field strength $B$ and the interaction $\boldsymbol{ V }$. In this notation then, the energy $ \Delta E \equiv \Delta E ( T, B, \boldsymbol{ V } ) $ required to excite the system is obtained from the expression $\Delta E = E_{ + } - E_{ - }$. Finally, substituting the left-hand side of Eq. (\ref{Eq:Classical_g_factor}) by  $\Delta E ( T, B, \boldsymbol{ V })$ and solving for $g_e$ and replacing $g_e \longmapsto g_{\text{eff}}$ yields Eq. (\ref{gEffective}).
The purpose of this article is to demonstrate that nuclear vibrations coupled to the electronic structure may have a significant influence on the $g$-factor $g_e$. To this end, we consider the electron-phonon self-energy in the Hartree-Fock approximation, providing an electron-phonon induced shift of the electronic energy levels (Hartree) and contribution to the exchange (Fock). The second order contribution  considered in this article is partitioned according to
\begin{align}
	{ \boldsymbol{ \Sigma } }^{(2)} ( \mathrm{i} \omega_n ) =& 
		{\boldsymbol{ \Sigma }}^\text{H} + {\boldsymbol{ \Sigma }}^\text{F}( \mathrm{i} \omega_n )
	.
\end{align}
The Hartree contribution is given by
\begin{align}\label{Eq:Hartree}
\boldsymbol{ \Sigma }^\text{H} ( \mathrm{i} \omega_n ) =
	\boldsymbol{ V }
	\mathscr{D}_0 ( \mathrm{i} \omega_m = 0 )
	\operatorname{Tr}
		\boldsymbol{ V } \boldsymbol{ C }
	,
\end{align}
where the non-interacting phononic Green-function $\mathscr{D}_0 ( \mathrm{i} \omega_m = 0 ) = - 2 / \Omega $ denotes the phonon-propagator evaluated at zero. Writing $n_{F}$ and $n_B$ for the Fermi-Dirac and Bose-Einstein distribution functions, respectively, one has the diagonal matrix $\boldsymbol{ C } = \operatorname{diag} ( n_F(\omega_+),n_F(\omega_-),n_F(\omega_+),n_F(\omega_-) ) $, whereas the trace extends over both orbital and spin degrees of freedom. The second self-energy contribution $\boldsymbol{ \Sigma }^\text{F}$ is the electron-phonon exchange loop given by
\begin{align}\label{Eq:Fock}
{\boldsymbol{ \Sigma }}^\text{F}( \mathrm{i} \omega_n ) 
=
	{\bf V} \widetilde{\boldsymbol{ \Sigma }}( \mathrm{i} \omega_n ) {\bf V}^\dagger
\end{align}
where $ \widetilde{\boldsymbol{ \Sigma }} ( \mathrm{i} \omega_n ) = \operatorname{diag} ( L_+( \mathrm{i} \omega_n ), L_-( \mathrm{i} \omega_n ) , L_+( \mathrm{i} \omega_n ), L_-( \mathrm{i} \omega_n ) ) / \hslash $ is a diagonal matrix and
\begin{align}
L_s(\mathrm{i} \omega_n )=&
	\frac{ n_B ( \omega ) + n_F ( \omega_s )}{ \mathrm{i} \omega_n  - \omega_s + \omega }
	+
	\frac{ n_B ( \omega ) + 1 - n_F ( \omega_s) }{ \mathrm{i} \omega_n  - \omega_s - \omega }
	,
\label{eq-Ls}
\end{align}
for $s=\pm$.
%
\section{Results and Discussion} \label{Sec:drivenSSH}
\begin{figure}
  \includegraphics[width=\columnwidth]{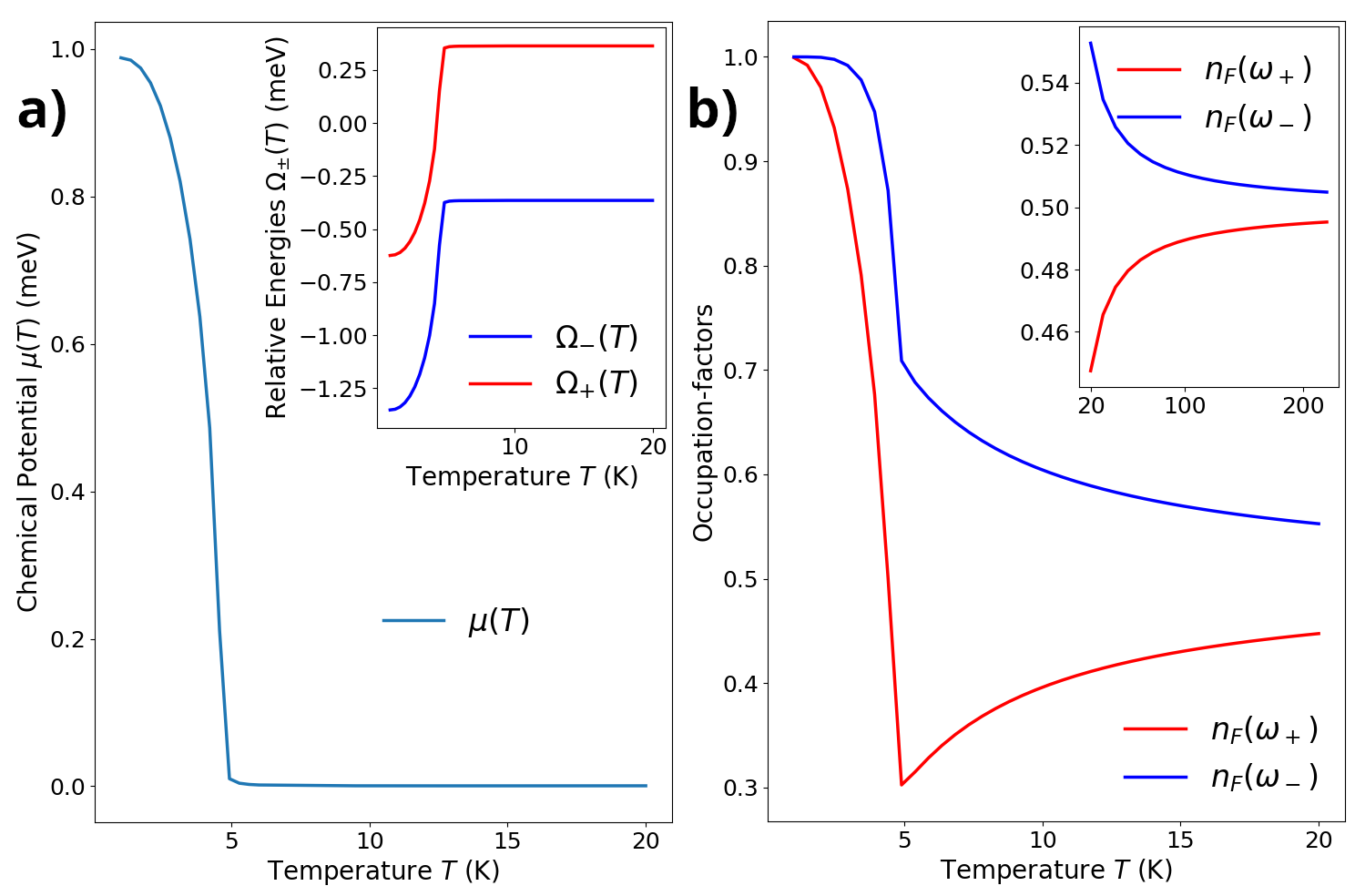}
  \caption{a) The chemical potential as a function of temperature, inset: The relative energies $\Omega_{\pm} ( T ) = \hbar \omega_{\pm} ( T ) = \pm \epsilon_z - \mu (T)$, b) The occupation-factors of the relative energies $n_F ( \omega_{\pm} )$, as a function of temperature for the interaction $ {\bf V} \equiv v_y \sigma_y $, in presence of a magnetic field of $B = 1$ T, and for an interaction strength of $v_y=20$ meV, and phonon-energy $\Omega = 200$ meV.}
  \label{fig:chemical_potential_and_occupations}
\end{figure}

\begin{figure}
  \centering
  \includegraphics[width=\columnwidth]{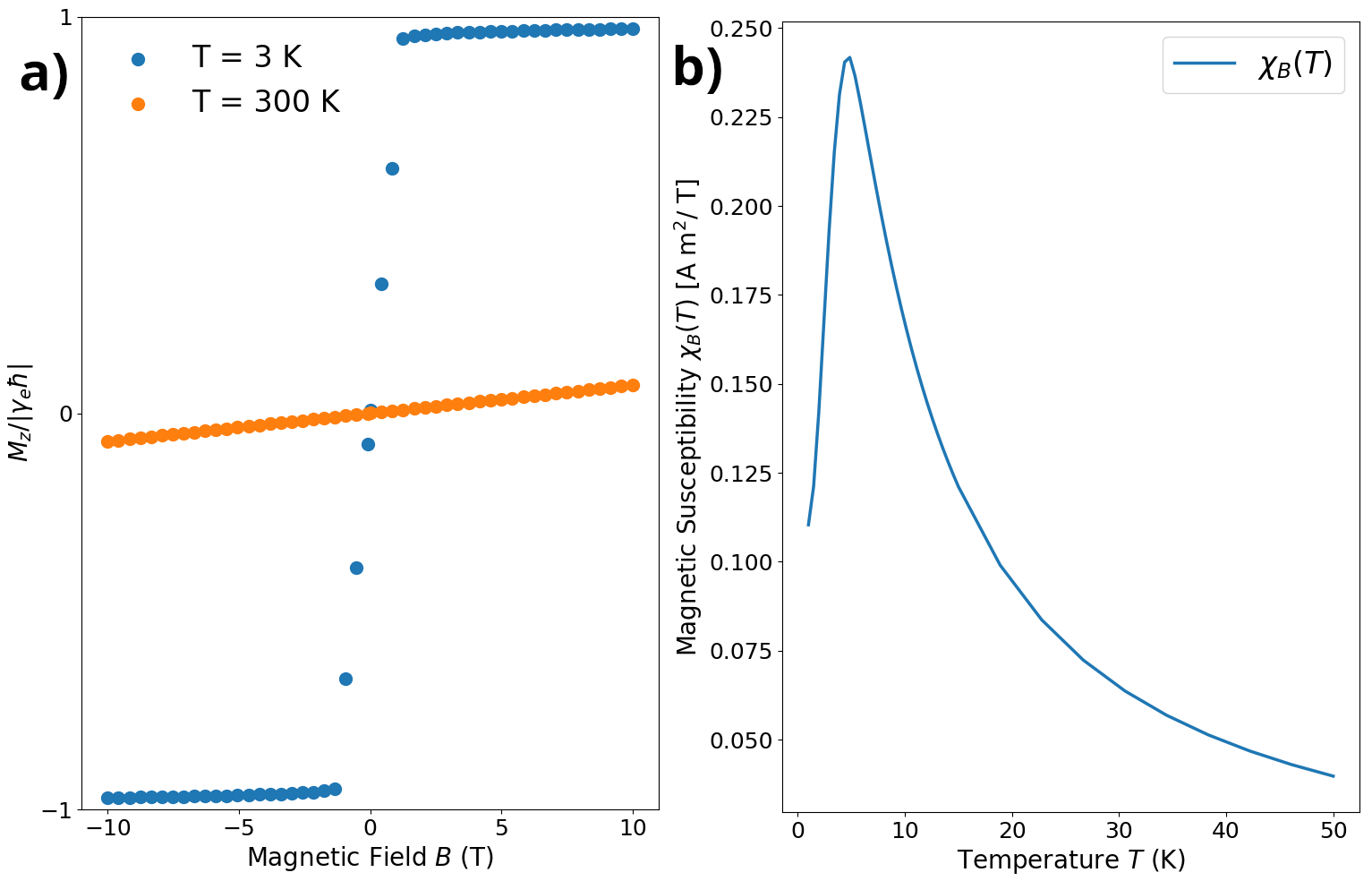}
  \caption{a) The magnetization above and below the critical temperature as a function of the external magnetic field, normalized w.r.t. the gyromagnetic ratio multiplied with Planck's constant $\gamma_e \hbar = - g_e \frac{ |e| \hbar }{2 m_e}$ with $g_e = 2.002$. b) The magnetic susceptibility as a function of temperature in presence of a magnetic field of $B = 1$ T. Both figures are computed for the interaction $ {\bf V} \equiv v_y \sigma_y $ and for an interaction strength of $v_y=20$ meV, and phonon-energy $\Omega = 200$ meV. The fact, that the magnetization is not constant until $B = 0$ is due to finite size effects of the regularization-parameter in Equation (\ref{Eq:Magnetization}). }
  \label{fig:Magnetic_Phase_Transition}
\end{figure}

As a minimal model, we consider the case in which the matrix element of the spin-orbit interaction that couples different spins to phonons is an imaginary number $ V_{ \substack{ 1 , 2 \\ + , - } } \equiv \langle \psi_{1, + } | H_{\text{soc-ph}} | \psi_{2 , - } \rangle = - \mathrm{i} v_y$, with $v_y$ being real-valued. Time reversal symmetry then implies $ V_{ \substack{ 1 , 2 \\ - , + } } \equiv \langle \psi_{1, - } | H_{\text{soc-ph}} | \psi_{2 , + } \rangle = + \mathrm{i} v_y$, and hence we obtain the interaction matrix $ {\bf V}_1 \equiv v_y \sigma_y $ whereas we used the decomposition of a $2\times2$-matrix with respect to the Pauli matrices.
Notice that in this case the Hartree-contribution of the self-energy vanishes, whereas the Fock-contribution acquires the simple form
\begin{align}
{\bf\Sigma}^\text{F}( \mathrm{i} \omega_n ) 
=
v_y^2 \widetilde{ \boldsymbol{ \Sigma }} ( \mathrm{i} \omega_n ) / \hslash
\end{align}
We consider typical field strengths $|B|\lesssim1$ T and optical phonon energies $ \Omega \lesssim 200$ meV \cite{105_Zdeb2024, 106_Chandrasekar2016, 107_Wu2020, 108_Wei2025, 109_Mohr2007, 110_Zhang2021, 111_Lucrezi2024}. We further restrict attention to a single phonon mode, which is a typical limiting case, suited for our model calculation \cite{112_Pfingsten2018, 110_Zhang2021, 09_Iaru2021}. The coupling strengths are taken to be of the order of a few $\text{meV}$ \cite{113_Giustino2017, 24_PONCE2016116, 25_Giustino2007, 115_Heid2017}. These values are system-dependent, however, as we shall see, the obtained effects are stable against typical variations of these values.
\begin{figure}[h]
  \includegraphics[width=\columnwidth]{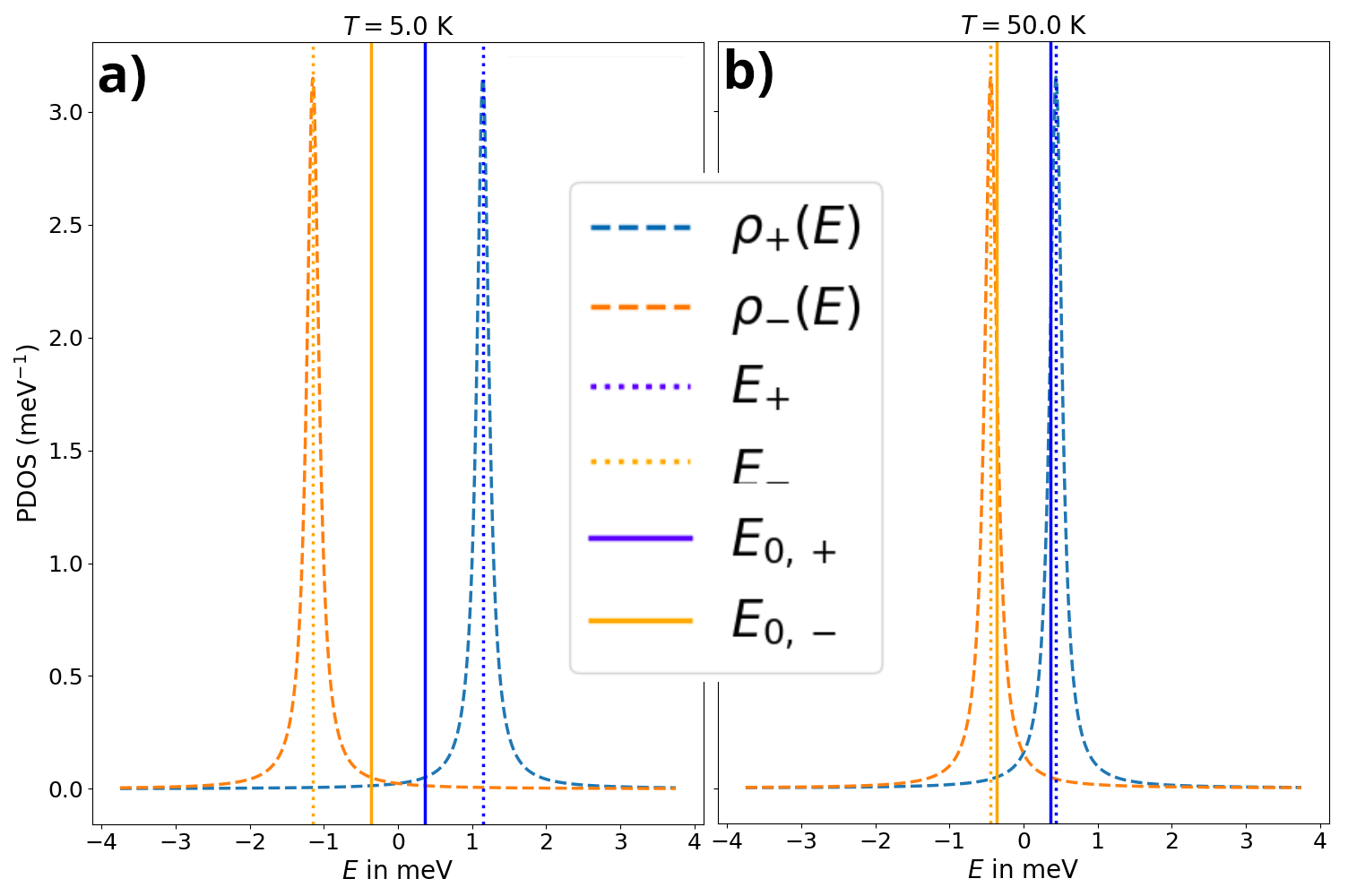}
  \caption{For $\sigma \in \{ \uparrow , \downarrow \}$ one has the average energies $E_{0, \sigma}$ (cf. Equation (\ref{eq:non_interacting_energies})) of the non-interacting electronic system in which interactions with phonons are absent (solid vertical lines), together with the centers of the peaks $E_{\sigma}$ (dotted vertical lines) of the partial density of states (PDOS) for the spin-states $\rho_{\sigma}$ (dashed lines) of the interacting system in which interactions with phonons are incorporated. The interaction is given as $ {\bf V} \equiv v_y \sigma_y $, with the interaction strength $v_y = 20$ meV and phonon-energy $\Omega = 200$ meV in a magnetic field of $B = 1$ T. a) At $T = 5$ K. b) At $T = 50 $ K.}
  \label{fig:DOS}
\end{figure}
As can be seen from Figure \ref{fig:DOS}, the partial densities of states $\rho_{ \pm} $ from Equation (\ref{eq:Spin-resolved-DOS}) have sharp peaks at energies $E_{ \pm }$, at which electronic states for the different spins are localized. The positions of these peaks depend on the temperature, interaction strength, and, as we shall see soon for low temperatures, on the magnetic field strength. After determining the electronic energies $E_{\sigma}$ for each temperature, we can use Equation (\ref{gEffective}) to determine the effective $g$-factor as a function of temperature.
\begin{figure}[h]
  \includegraphics[width=\columnwidth]{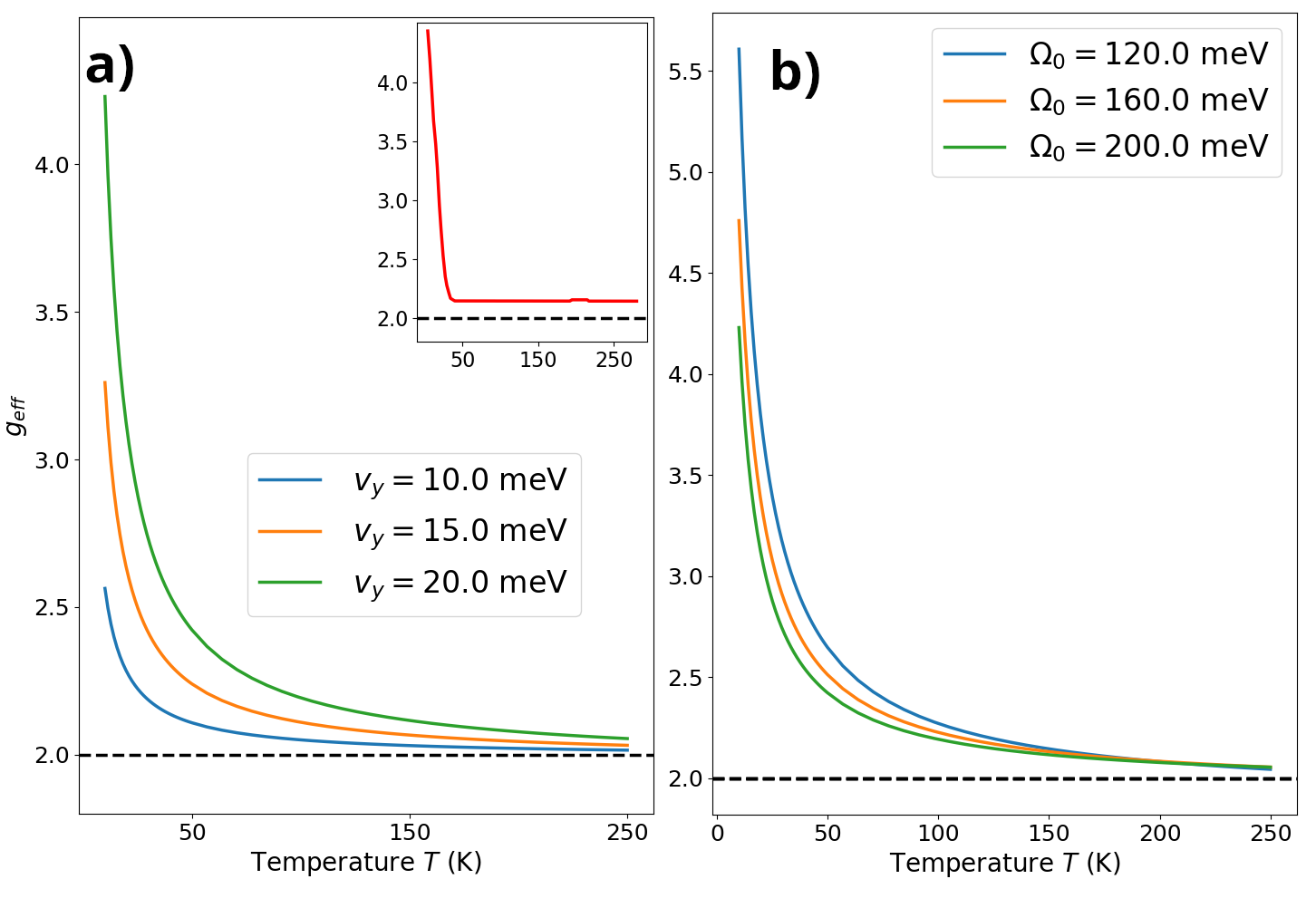}
  \caption{The effective $g$-factor as a function of temperature for the interaction $ {\bf V} \equiv v_y \sigma_y $, in presence of a magnetic field of $B = 1$ T. a) The effective $g$-factor for the interaction strengths $v_y=10$, $15$, and $20$ meV, and phonon-energy $\Omega = 200$ meV. The inset shows the comparison with experimental data taken from Ref. \citenum{15_gvsTref1}. b) The effective $g$-factor for the interaction strength $v_y=20$ meV, and phonon energies $\Omega=120$, 160, and 200 meV.}
  \label{fig:g_factors_case_1}
\end{figure}
The calculation of the effective g-factor as a function of temperature in Figure \ref{fig:g_factors_case_1} mirrors the traditional ESR approach. By substituting the excitation energies $\Delta E = E_+ - E_-$ that we derived from the partial density of states $\rho_{\pm}$ into the expression for $g$-factors in equation (\ref{gEffective}), we prove that our approach provides insights that are consistent with and are also expanding on classical ESR interpretations.
The effective $g$-factor acquires a temperature dependence imposed by the phononic contribution which leads to a monotonous decrease with increasing temperature in the given temperature-range of Figure \ref{fig:g_factors_case_1}. Additionally, the $g$-factor decreases strongly at low temperatures while the decrement is weak to absent for higher. It can also be noticed that this type of temperature dependence is universal with respect to the coupling strength $v_y$, Fig. \ref{fig:g_factors_case_1} a), the phonon energy $\Omega$, Fig. \ref{fig:g_factors_case_1} b). The experimental data, shown in the inset of Figure \ref{fig:g_factors_case_1}, was obtained from Ref. \citenum{15_gvsTref1} for the measured $g$-factor of polycrystalline $\text{CaCu}_3\text{Ti}_4\text{O}_{12}$. The authors attribute the sudden rise of the $g$-factor at low temperatures to a ferromagnetic-paramagnetic phase transition, whereas the ferromagnetic behaviour refers to the alignment of the spins inside the [111] Cu$^{2+}$-planes, and there is an antiferromagnetic ordering along the (111) direction \cite{130_Koitzsch2002}. Remarkably, our model successfully captures this behavior without the need for additional assumptions. This can be seen most directly, by looking at the abrupt change of the chemical potential and magnetic susceptibility as a function of temperature, at around $T \equiv T_{\text{c}} \approx 5 $ K in Figures \ref{fig:chemical_potential_and_occupations} a), \ref{fig:Magnetic_Phase_Transition} b). Further, one observes the usual magnetization-curves above and below the critical temperature $T_{\text{c}}$ in Figure \ref{fig:Magnetic_Phase_Transition} a), that are typical for ferro- and paramagnetism. The phonon energies in $\text{CaCu}_3\text{Ti}_4\text{O}_{12}$ have been reported to reach up to approximately 100 meV at the $\Gamma$-point, cf. ref. \cite{151_PhysRevB.65.214112, 152_PhysRevB.77.045131}, placing the system in a regime that is compatible with our model. While the precise electron-phonon coupling strength depends on the full phonon dispersion and coupling matrix elements, which are challenging to compute for magnetic systems with large unit cells, the qualitative agreement between the model and experimental data supports the relevance of strong spin-dependent electron-phonon interactions in this material. The observed temperature-dependent rise in the $g$-factor can be naturally explained by the effective spin-phonon coupling described in our framework. While the difference in $g$-factors at high temperatures arises from the omission of phonon-independent spin-orbit interactions to highlight the emergence of the temperature dependence, our model provides an excellent reproduction of the experimentally observed phase transition behavior, as shown in the inset of Fig. \ref{fig:g_factors_case_1} a) \cite{15_gvsTref1}. 
Similar results as the one referenced above were found by others \cite{18_gvsTref5, 14_gvsTref2, 13_gvsTref3, 45_Sichelschmidt2007-gc, 15_gvsTref1, 17_gvsTref4}, for instance in magnetite, $\text{Fe}_3 \text{O}_4$, nanoparticles in \cite{14_gvsTref2}. In this experiment, the temperature dependence of the ESR spectrum of the nanoparticles was studied. Whereas the authors used an equation as $g_\text{eff}(T) = a / T^2 + b / T + c$ in order to fit the observed spectrum. Such a function is compatible with our results in the relevant experimental temperature regime. Furthermore, our model has the advantage that it can even be extended to ab initio calculations, by explicitly computing the coupling-constants and, hence, \emph{predicting} the temperature dependence of the $g$-factor. 
The qualitative behaviour of the $g$-factor for optical phonons approximately independent of the concrete value of the phonon energy $\Omega$, Fig. \ref{fig:g_factors_case_1} b), particularly for low temperatures. The same holds true for small changes in the coupling strength $v_y$. The qualitative behaviour is always the same and is in excellent agreement with several experimental studies.
\begin{figure}
  \includegraphics[width=\columnwidth]{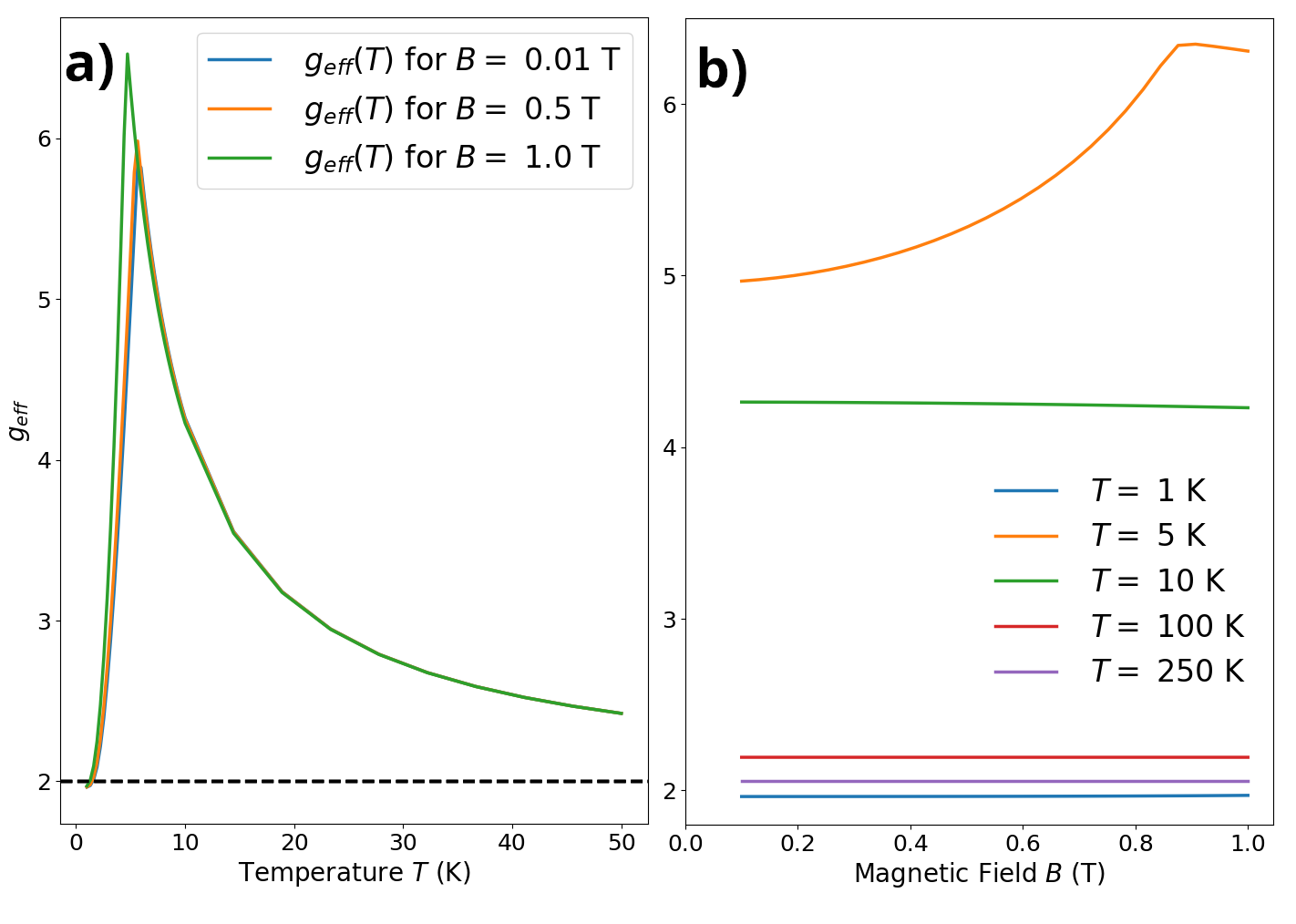}
  \caption{The effective $g$-factor as a function of temperature for the interaction $ {\bf V} \equiv v_y \sigma_y $, for for the interaction strength $ v_y = 20 $ meV and phonon energy $\Omega = 200 $ meV. a) The effective $g$-factor for magnetic fields of $B = 0.01, 0.5, 1.0 \,$ T at low temperatures. b) As in a) for a larger range of temperatures.}
  \label{fig:g_factors_case_2}
\end{figure}
In addition to retaining the temperature dependence of the $g$-factor obtained experimentally, we also consider temperatures below the ones reported from the measurements. Our prediction for temperatures $T<5$ K, is a non-monotonous behavior of the $g$-factor. In Fig.~\ref{fig:g_factors_case_2}~a),~b) we have plotted the calculated $g$-factor for temperatures below $T<20$~K for different magnetic field strengths, showing a pronounced peak structure of the $g$-factor at low temperatures. At temperatures below this peak the $g$-factor approaches the non-interacting value of $g \approx 2$ as $T\rightarrow0$. This behavior can be attributed to the unconventional temperature dependence of the phonon occupation number in the interacting system. In the non-interacting case, the phonon occupation number $\langle b^{\dagger} b \rangle_0 = n_B(\omega)$ vanishes exponentially as $T \to 0$. However, in the presence of spin-dependent electron-phonon interactions, the phonon sector becomes entangled with the electronic degrees of freedom. As a result the standard Bose-Einstein behavior of the phononic occupation number is modified. Moreover, in Fig. \ref{fig:g_factors_case_2} one sees that the $g$-factor has a dependence on the magnetic-field at low temperatures, albeit not strong. To our knowledge, the non-monotonicity of the $g$-factor has not been reported, and the magnetic field dependence only in heavy-fermion compounds, but not for system exhibiting non-negligible spin orbit interactions. As clearly seen in Fig. \ref{fig:g_factors_case_2} a), the $g$-factor peaks at a magnetic field dependent temperature, and its origin can be derived back to the electron-phonon interactions. The observed sharp increase in the chemical potential $\mu ( T )$ as observed in Figure \ref{fig:chemical_potential_and_occupations} a) and the kink in the effective $g_{ \text{eff} } (T) $ near $5$ K as observed in Figure \ref{fig:g_factors_case_2} a) highlight a temperature-dependent transition in the electron-phonon coupling regime. This behavior suggests a crossover from a weakly coupled regime at higher temperatures to a strongly coupled regime at lower temperatures. The chemical potential exhibits a sharp change in the first derivative at low temperatures, signaling a magnetic phase transition in accordance with the Ehrenfest classification  \cite{120_Nolting2012-xl, 122_Schwabl2006-sb, 123_Fliebach2018-nn}. The magnetic field dependence of the $g$-factor is only visible at low-temperatures. The explanation for this behaviour can be traced back to the interpretation of the mathematical expressions: The Hartree-contribution to the electron-phonon self-energy in Eq. \eqref{Eq:Hartree} vanishes for our choice of interaction. The Fock-contribution to the electron-phonon self-energy in Eq. \eqref{eq-Ls} comprises two components, of which the first and second are referred to as the emission and absorption processes, respectively.
For the temperatures considered in this paper, the average phonon occupation is negligible, hence, the spectral weights of the self-energy are solely determined by the electronic occupation.
It is important to notice that the unperturbed occupation factors $ n_F ( \omega_{\pm} ) $ for these states are in the interacting case $n_F(\omega_+) = n_F(\omega_-) \approx 1$ at $T=0$ K, see Fig. \ref{fig:chemical_potential_and_occupations} b). Hence, for low temperatures, the absorption and emission processes are both active for the up- and down-states, with weight close to unity. By increasing the temperature, the occupation-factors $ n_F ( \omega_{\pm} ) $ divert from their common value at $T=0$ K, meaning that the spin-dependent electron-phonon processes contribute unequally to the relative energies of the two spin-states. One of the spin states loses occupation faster than the other spin state. In the high-temperature-limit, as can be seen in the inset of Figure \ref{fig:chemical_potential_and_occupations} b), both occupation-factors approach 1/2, such that spin-up and spin-down states experience the same energy-shifts, which leads to a $g$-factor of 2. Hence, both electron-phonon processes, emission and absorption, are active for both spin-states with different intensities at different temperature-scales such that the real part of the self-energy generates several competing energy shifts. Therefore, the effective Zeeman-splitting that results from the combination of the applied magnetic field and the contribution from the self-energy, is not constant as function of temperature. The combined effect of phonon-creation and phonon-annihilation creates a relative energy shift in the electronic density of states, and hence renormalizes the Zeeman-energies.
The temperature-behaviour of the $g$-factor may also be comprehended from the classical picture of vibrations, which \emph{freeze out} at very low temperatures. The dramatic increase in the $g$-factor at intermediate temperatures is due to the electron-phonon interactions. Specifically, the temperature-dependence of the chemical potential changes the relative energies $ \hbar \omega_{\pm}$ in a subtle way. As the temperature rises, phonons become thermally excited, while simultaneously the occupation-factors $n_F( \omega_{ \pm} )$ differ for the different spin-states, which leads to shifts in the $g$-factor. At higher temperatures there is a damping of the electron-phonon interactions, due to the asymptotics of the occupation-factors. The system reaches a regime, where the phonon occupation number is large but the coupling effects on the $g$-factor asymptotically diminish. The exact value and temperature-dependence of the $g$-factor depends on the specifics of the materials coupling coefficients.
\begin{figure}[h]
  \centering
  \includegraphics[width=\columnwidth]{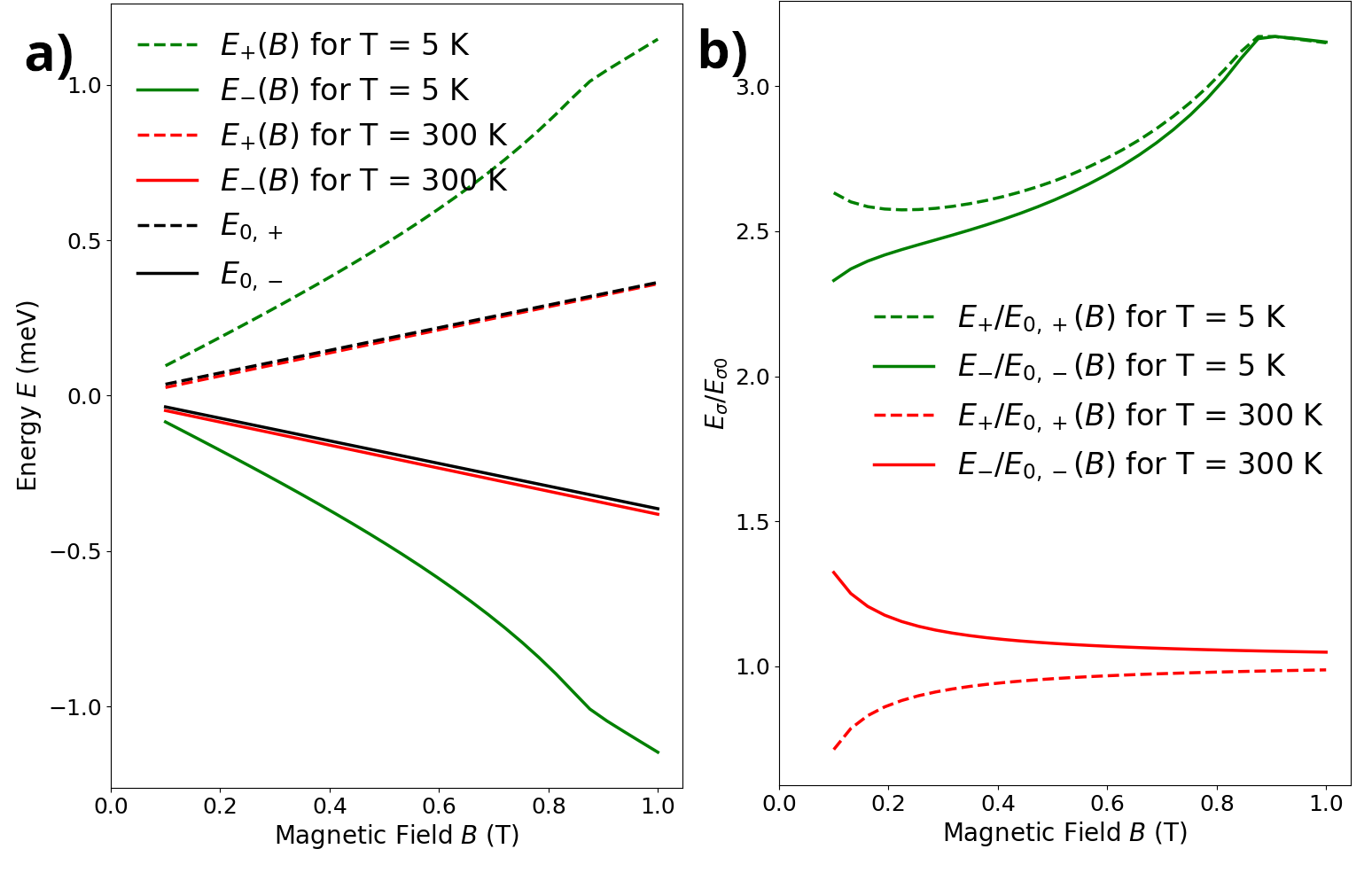}
  \caption{The spin energy levels as functions of the magnetic field strength $B$. a) The energy levels $E_\sigma$, $\sigma=\uparrow,\downarrow$, for the temperatures $T=5$ and 350 K. b) The normalized energy level $(E_\sigma/E_{0\sigma})$ for the temperatures $T=5$, 100, 200, and 300 K. The coupling strength $v_y = 20 $ meV, while $\Omega = 200 $ meV as in Fig. \ref{fig:g_factors_case_1} a)}
  \label{fig:energy_levels}
\end{figure}
In Fig. \ref{fig:energy_levels} a) we plot the electronic energy spectrum as function of the magnetic field strength $B$, for the temperatures $T=5$ and 350 K. The bare energies $E_{0\sigma}$, that is, in absence of phonons, are linear in the magnetic field and converge in the origin, as expected. By contrast, the phonons add a temperature-dependent shift in the energies and the energy levels are also equipped with non-linear dependencies on the magnetic fields, which is stressed in Fig. \ref{fig:energy_levels} b) where the ratios $E_\sigma/E_{0\sigma}$ are displayed.
%
%
\section{Conclusion} \label{Sec:summary}
%
In summary, we have proposed a mechanism for the previously observed temperature dependencies of the $g$-factor. By inclusion of spin-dependent electron-phonon interactions, we theoretically show that the $g$-factor acquires strongly temperature dependent properties which are universal under variation of the electron-phonon coupling strength, the phonon energies, and the strength of the external magnetic field. This reinterpretes the role of phonons in ESR-spectroscopy beyond the traditional viewpoint of them being a mere relaxation-channel. The simultaneous change in $\mu (T) $ and $ g_{ \text{eff} } (T) $ is indicative of a coupled evolution of the electronic system with the phononic background. This interplay becomes particularly interesting at low temperatures, where phonon-mediated effects reorganize the electronic energy spectrum. At these temperatures, the models implications could be tested by measuring the non-monotonicity of the $g$-factor.
The discontinuity in the first derivative of the chemical potential and the magnetic susceptibility as functions of temperature, and the magnetization as a function of the external magnetic field provide robust thermodynamic markers for a magnetic phase transition, directly connecting theoretical predictions with experimental observations. We also find that the energy spectrum of electron-phonon coupled system varies non-linearly with the strength of the applied magnetic field close to the phase transition. This is another aspect which can be experimentally tested. 
This work thus bridges traditional ESR methods and advanced models by incorporating electron-phonon interactions, demonstrating the flexibility of using the density of states to determine excitation spectra and $g$-factors. The theoretical predictions presented in this work can, in principle, be tested experimentally. Specifically, we suggest applying electron spin resonance (ESR) spectroscopy under conditions similar to those used in Ref. \cite{41_31_Referee_Article},
with frequencies between 9 and 360 Ghz, magnetic fields ranging from 0.2 to 14 T, and temperatures in the range 2-25 K, to the compound CaCu$_3$Ti$_4$O$_{12}$ \cite{15_gvsTref1}. This setup appears well-suited for probing the phenomena described in the manuscript. Lastly, our framework can also be used to study system with electron-electron interactions, such as heavy-fermion compounds.
We hope our findings might offer new perspectives on the current interpretations of the many spin resonance measurements conducted daily.
\subsection*{Data Availability}
The data for the plots of this study are available at \cite{kalhofer_2025_14930497}.
\subsection*{Acknowledgments}
The author thanks Jonas Fransson for his valuable discussions and guidance, 
which greatly contributed to the conception and development of this work. 
The author also expresses gratitude to Annica Black-Schaffer for insightful feedback and support. 
Special thanks go to Lucas Baldo and Rodrigo Arouca for their invaluable feedback and for proofreading drafts of the manuscript, 
as well as for the many fruitful discussions that shaped the ideas presented here. Further thanks goes to Arnob Kumar Ghosh, for helping with the submission process.
The computations were enabled by resources provided by the National Academic Infrastructure for Supercomputing in Sweden (NAISS) at \url{https://www.naiss.se/} partially funded by the Swedish Research Council through grant agreement no. 2022-06725.

\bibliographystyle{apsrev4-2mod}
\bibliography{ref.bib}
\end{document}